# On-chip processing of optical orbital angular momentum

*Charles Chen[1]*

*1 State Key Laboratory for Mesoscopic Physics & Department of Physics, Collaborative Innovation Center of Quantum Matter & Frontiers Science Center for Nano-optoelectronics, Peking University, Beijing 100871, P. R. China*

# Abstract

Orbital angular momentum is an important concept in optics, thus numerous researches explore the principles and applications of light beams with orbital angular momentum. This type of light beam is also called vortex beam, whose inherent infinite-dimensional feature has broad prospect in many fields like optical communications and optical computing. However, there are still a lot of problems in the propagation and processing of on-chip vortex beams. Here, with inverse design method, we design four cross-shaped waveguide structures which can support vortex beams with topological charge $l$ = 1, 2, 3, 4 respectively. The four vortex modes are momentum matched so the coupling between every two waveguides can happen, achieving the conversion of orbital angular momentum. The longest width of each waveguide is about 2 μm, so the cross-shaped waveguides are able to be integrated on a photonic chip. All the cross-shaped waveguides are silicon-on-insulator waveguide, and such silicon-based platform has the advantage of being compatible with complementary metal oxide semiconductor technology. Moreover, in order to demonstrate the application prospect, we construct an XOR logic gate with the conversion of orbital angular momentum. Therefore, our work provides a new idea for the manipulation of orbital angular momentum, expands the applications of on-chip vortex beams, and shows great potential.

# 1 Introduction

In the field of modern optics, orbital angular momentum (OAM) has emerged as an important concept with far-reaching implications[1-4]. OAM light beams, characterized by a helical phase front and an azimuthal phase factor of $e^{il\varphi}$, where $l$ is the topological charge and $\varphi$ is the azimuthal angle, possess an inherently infinite dimensional property[5]. This unique feature has spurred intense research efforts, as it holds great promise for many applications.

In order to fully utilize such an advantage, many fields such as optical computing[6, 7], multiplexing[8, 9], communication[10, 11], quantum technology[12-14] and micro-manipulating[15, 16] have done many research around OAM. As a result, on-chip integration has become an inevitable path for OAM-supporting devices. This integration is crucial for miniaturization, compatibility with existing photonic integrated circuits, and achieving adaptable technological advancements. Previous works have achieved the generation of OAM light beams on-chip[17], and many waveguide structures for guiding OAM light beams have been proposed[18, 19], such as silicon waveguides with cross-shaped cross-section[20]. However, how to manipulate the topological charge of OAM light on complementary metal oxide semiconductor (CMOS) compatible platform to utilize this new degree of freedom remains a challenging problem. To address the challenges in the process of miniaturization and integration, inverse design has emerged as a powerful tool[21]. Inverse design, unlike traditional forward design, starts from the desired functionality and work backward to determine the optimal structure. Compared with forward design, inverse design has obvious advantages[22, 23]. It can break through the limitations of forward design, start from the target function, and does not rely on prior physical knowledge. By means of algorithms, it can efficiently search in the complex design space, quickly and accurately find optimization solutions, and effectively overcome the local minimum problem.

In this work, we also adopt the concept of inverse design and utilize the genetic algorithm (GA) to optimize the cross-shaped waveguide (CSW) structures. As a result, we obtain two CSW structures, which can respectively support the $OAM_{l=1}$ mode and $OAM_{l=2}$ mode with degenerate effective refractive indexes $n_{eff}$. Subsequently, based on the coupling mode theory (CMT), we successfully achieve the mutual conversion between OAM light beams with two different topological charges, realizing the on-chip manipulation of OAM light. Fig. 1 shows the schematic diagram of our designed system. Besides, we also verify the scalability of our work. We further optimize CSW structures which can support degenerate $OAM_{l=3}$ mode and $OAM_{l=4}$ mode, and these two modes can couple with the OAM$_{l=1}$ mode and OAM$_{l=2}$ mode as well. Eventually, we construct an XOR gate by converting the on-chip OAM modes, which provide a brief demonstration of its potential applications.

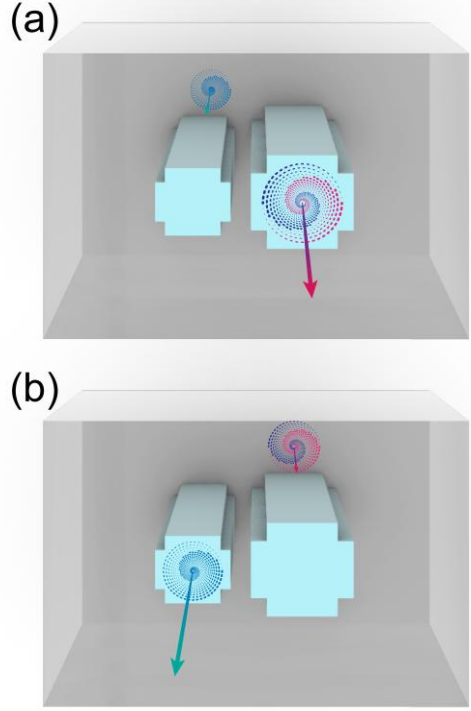

**Fig. 1** Schematic diagram for conversion of OAM. **a** OAM$_{l=1}$ mode is input and converted to OAM$_{l=2}$ mode. **b** OAM$_{l=2}$ mode is input and converted to OAM$_{l=1}$ mode.

## 2 Inverse design of CSW parameters using GA

The most common OAM modes are Laguerre-Gaussian (LG) modes, but they are not the eigenmodes supported by waveguides in most cases. Previous works have demonstrated that specially designed cross-shaped silicon-on-insulator (SOI) waveguides can transmit LG modes with different topological charge number $l$[20]. Our goal is to achieve on-chip OAM conversion by coupling OAM light beams of different orders, thereby controlling their topological charges. According to the CMT, the coupled modes need to be momentum matched, meaning that the effective refractive indexes $n_{eff}$ of the two modes should be as approximate as possible. Therefore, it is necessary to redesign the parameters of the CSW structure. With the final goal clarified, we can use algorithms to inversely design the parameters of the CSW.

In this work, we employ the GA for inverse design. The GA is an optimization algorithm based on the principles of biological evolution, simulating the "survival of the fittest" mechanism to search for the optimal solution in the solution space[24]. The GA first initializes a population, which consists of multiple individuals. Each individual has its own chromosome, which is obtained by encoding the parameters to be optimized. Next, we design a fitness function to evaluate the quality of each individual in the population. Individuals with low fitness are eliminated, while those with high fitness are retained. Through crossover and mutation, a new generation of the population is generated, completing one evolutionary cycle. Finally, iterative optimization is performed by repeating selection, crossover, and mutation until the maximum number of iterations is reached or a satisfactory solution is found.

We consider the cross-section of the CSW as two overlapping, perpendicular rectangles. Thus, the shape of each CSW is determined by four parameters, that is, the length and width of the two rectangles. Since the coupling behavior occurs between two waveguides, we need to design two structures: CSW-1 and CSW-2, requiring optimization of eight parameters (the length and width of four rectangles). These eight parameters are encoded into chromosomes, and iterative optimization is performed. Since LG modes can be decomposed (decomposition of LG modes as shown in Fig. S1), according to the CMT, the $HG_{01}$ and $HG_{10}$ modes decomposed from the $LG_{01}$ mode need to achieve momentum matching with the $LG_{02}^o$ and $LG_{02}^e$ modes decomposed from the $LG_{02}$ mode. The goal of iterative optimization is to make the $n_{eff}$ of the $HG_{01}$, $HG_{10}$ modes supported by CSW-1 as close as possible to the $n_{eff}$ of the $LG_{02}^o$, $LG_{02}^e$ modes supported by CSW-2. Therefore, the fitness function is defined as:

$$f_1 = \frac{1}{\max(n_1, n_2, n_3, n_4) - \min(n_1, n_2, n_3, n_4)}, \qquad (1)$$

here, $n_1$ represents the $n_{eff}$ of the $HG_{01}$ mode, $n_2$ represents the $n_{eff}$ of the $HG_{10}$ mode, $n_3$ represents the $n_{eff}$ of the $LG_{02}^o$ mode, and $n_4$ represents the $n_{eff}$ of the $LG_{02}^e$ mode. Previous studies have shown that when the difference of $n_{eff}$ between two modes is on the order of $10^{-4}$, the two modes can be considered effectively separated[25, 26]. Therefore, we set the degeneracy criterion as the difference of $n_{eff}$ of two modes below $10^{-4}$, which means the fitness function $f_1 > 10^4$.

We set the population size to 50, and after 60 generations of iterations, we obtain the iterative process, which is shown in Fig. 2(a). The individual with the highest fitness function value achieves $f = 1.52 \times 10^4$, corresponding to the $n_{eff}$ difference of $6.6 \times 10^{-5}$. At this point, the cross-sections of CSW-1 and CSW-2, as well as the optical field intensity and phase distributions of the OAM$_{l=1}$ mode and OAM$_{l=2}$ mode, are also shown in Fig. 2.

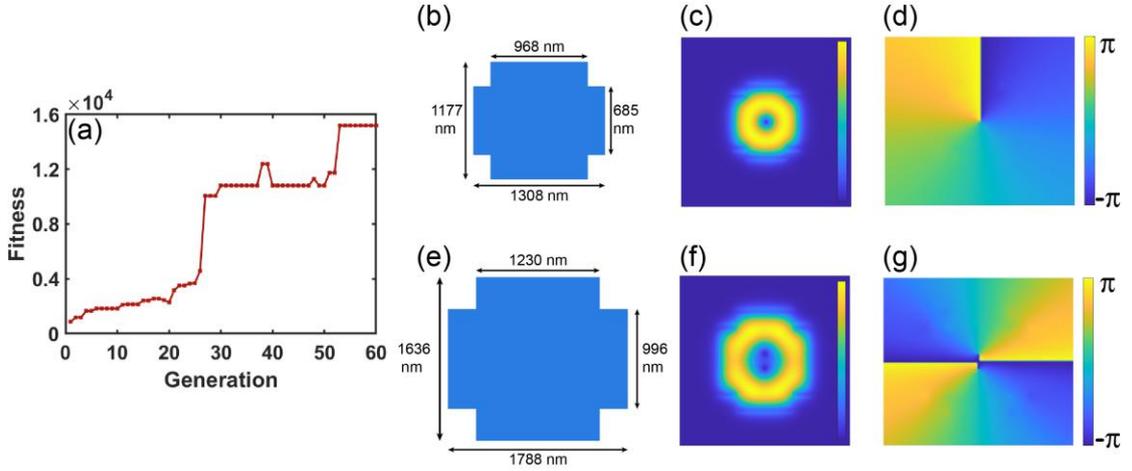

**Fig. 2** The CSW-1 and CSW-2 designed in reverse and the OAM$_{l=1}$ mode and OAM$_{l=2}$ mode that they support. **a** The iterative process obtained when optimizing CSW-1 and CSW-2 using Eq. (1) as the fitness function. **b** Cross-section of CSW-1. **c** Intensity distribution and **d** phase distribution of OAM$_{l=1}$ mode which is guided in CSW-1. **e** Cross-section of CSW-2. **f** Intensity distribution and **g** phase distribution of OAM$_{l=2}$ mode which is guided in CSW-2.

## 3 Converting OAM by mode coupling

Through inverse design optimization, we obtain two CSW structures. The OAM$_{l=1}$ mode supported by CSW-1 and the OAM$_{l=2}$ mode supported by CSW-2 satisfy the momentum matching condition. Therefore, by placing the two waveguides side by side with an appropriate gap, coupling between these two modes can be achieved[17, 27]. Simulations reveal that there will be effective coupling when the gap is set as 252 nm, just like Fig. 3(a) shows.

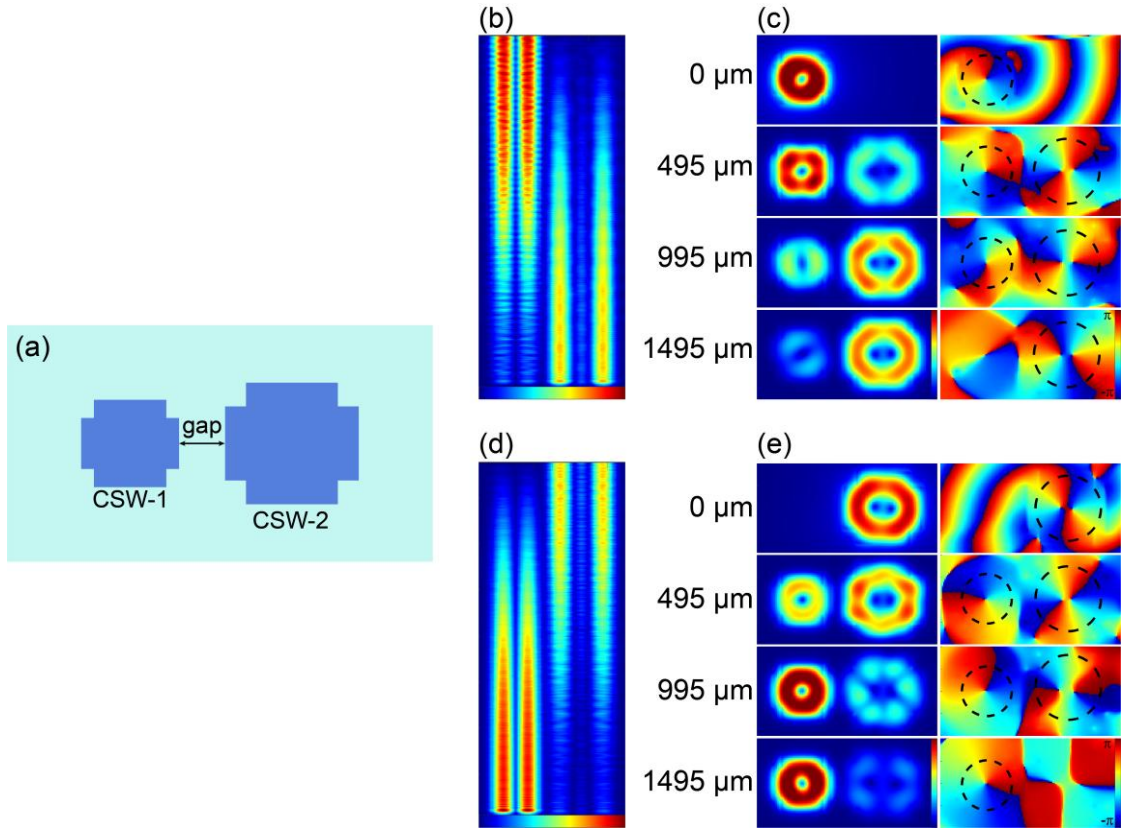

**Fig. 3** Coupling situations between $OAM_{l=1}$ mode guided in CSW-1 and $OAM_{l=2}$ mode guided in CSW-2. **a** Cross-section of the coupler. **b, d** Power flow distributions of the coupling from **b** $OAM_{l=1}$ mode to $OAM_{l=2}$ mode and **d** $OAM_{l=2}$ mode to $OAM_{l=1}$ mode. **c, e** Intensity and phase distributions in different coupling lengths of the coupling from **c** $OAM_{l=1}$ mode to $OAM_{l=2}$ mode and **e** $OAM_{l=2}$ mode to $OAM_{l=1}$ mode.

After determining the gap, we simulate the transmission of the coupled waveguide system. Generally, the optical power in the coupled waveguides exhibits periodic variations, and the period is defined as the coupling length. The transmission characteristics of OAM$_{l=1}$ mode coupling into OAM$_{l=2}$ mode are shown in Fig. 3(b)-(c), while the reverse process (OAM$_{l=2}$ coupling into OAM$_{l=1}$) is depicted in Fig. 3(d)-(e). Both processes exhibit a coupling length of about 1500 μm.

## 4 Discussion

### 4.1 Scalability

The approach of optimizing waveguide structures using the GA demonstrates strong scalability. Building on the foundation mentioned above, we further optimize CSW-3 and CSW-4 which are capable of supporting OAM$_{l=3}$ and OAM$_{l=4}$ modes. These higher-order OAM modes can also couple with the OAM$_{l=1}$ and OAM$_{l=2}$ modes which are guided in the CSW-1 and CSW-2.

We set the median $n_{eff}$ of the OAM$_{l=1}$ mode (supported by CSW-1) and the OAM$_{l=2}$ mode (supported by CSW-2) as $n_0$,

$$n_0 = \frac{n_1 + n_2 + n_3 + n_4}{4}. \tag{2}$$

Let the $n_{eff}$ of $LG_{03}^o$ and $LG_{03}^e$ be $n_5$ and $n_6$ respectively, then the new fitness function can be set as:

$$f_2 = \frac{1}{\max(n_5, n_6) - n_0}. \tag{3}$$

The CSW-3 optimized in this way supports the OAM$_{l=3}$ mode, which is degenerate with the OAM$_{l=1}$ mode in CSW-1 and the OAM$_{l=2}$ mode in CSW-2, thus enabling the mutual coupling. Similarly, we have also optimized the CSW-4, which supports the OAM$_{l=4}$ mode that is degenerate with the OAM$_{l=1}$, OAM$_{l=2}$ and OAM$_{l=3}$ modes. Fig. 4(a) and Fig. 4(d) show their iterative process. The intensity and phase distributions of the OAM$_{l=3}$ and OAM$_{l=4}$ modes are shown in Fig. 4(b)-(c) and Fig. 4(e)-(f).

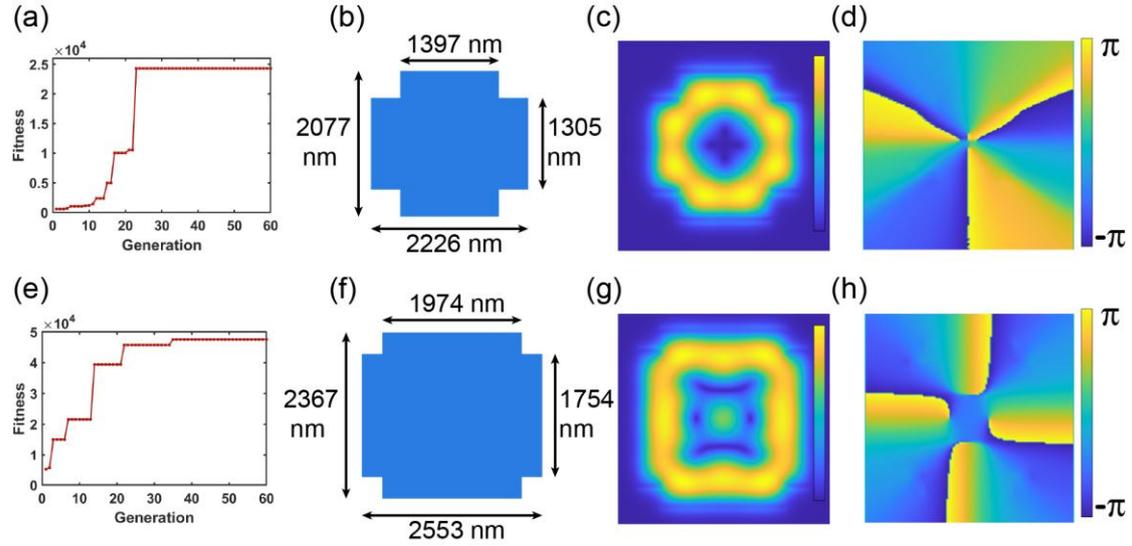

**Fig. 4** Optimizations and the supported OAM modes of **a-d** CSW-3 and **e-h** CSW-4. **a** The iterative process for optimization of CSW-3. **b** Cross-section of CSW-3. **c** Intensity and **d** phase distributions of OAM$_{l=3}$ mode guided in CSW-3. **e** The iterative process for optimization of CSW-4. **f** Cross-section of CSW-4. **g** Intensity and **h** phase distributions of OAM$_{l=4}$ mode guided in CSW-4.

The OAM modes ($l$ = 1, 2, 3, 4) supported by CSW-1, 2, 3, 4 are all mutually degenerate, therefore these modes can couple with each other. Fig. S2 illustrates the eigenmodes when these modes are coupled.

### 4.2 Applications

Utilizing mode coupling to manipulate the topological charge of on-chip OAM modes has numerous application values, such as logic gates. In the past, many studies have used coupler to construct logic gates[28-31]. The simplest logic gates, like AND gates and NOT gates, can be constructed by simply controlling the coupling length. In this work, we also introduce the degree of freedom of polarization and construct an XOR logic gate, which demonstrates the application value of using mode coupling to manipulate the OAM topological charge. We name this XOR gate OAM-XOR gate. All the degenerate modes mentioned above are the common TE modes in waveguides.

For waveguide modes of the same order, the $n_{eff}$ of TM modes and TE modes are not exactly the same. Therefore, the TM modes of OAM$_{l=1}$ and OAM$_{l=2}$ supported by CSW-1 and CSW-2 do not meet the momentum matching condition and thus will not couple with each other.

We propose a novel encoding mechanism using two optical degrees of freedom. We set the polarization state as input A, encoding the TM mode as 0 and the TE mode as 1. We set the OAM topological charge number as input B, encoding OAM$_{l=1}$ as 0 and OAM$_{l=2}$ as 1. At the output end, we also encode OAM$_{l=1}$ as 0 and OAM$_{l=2}$ as 1. Through the coupler we design, we can obtain Table 1, which is exactly the truth table of an XOR gate. The corresponding intensity and phase distributions of input modes and output modes are shown in Fig. 5.

Table 1 Truth table of the OAM-XOR gate

| Input A (polarization) | Input B (topological charge) | Output (topological charge) |
| --- | --- | --- |
| 0 (TM) | 0 ($l$=1) | 0 ($l$=1) |
| 0 | 1 ($l$=2) | 1 ($l$=2) |
| 1 (TE) | 0 | 1 |
| 1 | 1 | 0 |

| Polarization | Input | Output |
|---|---|---|
| TE | | |
| TM | | |
| TE | | |
| TM | | |

**Fig. 5** The input modes and the corresponding output modes of OAM-XOR gate. When the polarization of input mode is TE polarization, if we input $OAM_{l=1}$ mode, $OAM_{l=2}$ mode will be output, and vice versa. When the polarization of input mode is TM polarization, the output mode will be the same as the input mode.

## 5 Methods

All the simulations are finished with *Lumerical MODE* and *Lumerical FDTD*. The CSWs are all made of Silicon, and they are surrounded by $SiO_2$. The wavelength is set as 1550 nm. The eigenmodes are simulated by *Lumerical MODE*, which are painted in *Matlab* colormap *Parula*. The transmission situations are simulated by *Lumerical FDTD*, which are painted in *Matlab* colormap *jet*. The fabrication process of the samples is extremely complex, as shown in Fig. S3. We believe that this approach holds great promise for success.

## 6 Conclusions

In this work, we employ the GA to inversely design two OAM-supporting waveguide structures, denoted as CSW-1 and CSW-2. The $OAM_{l=1}$ mode in CSW-1 and the $OAM_{l=2}$ mode in CSW-2 are momentum matched, which enable mutual coupling and therefore achieve on-chip OAM conversion. To demonstrate scalability, we further design CSW-3 and CSW-4, which support $OAM_{l=3}$ and $OAM_{l=4}$ modes, respectively.

All four modes (OAM$_{l=1}$ to OAM$_{l=4}$) exhibit degenerate $n_{eff}$, allowing arbitrary pairwise coupling. Finally, to illustrate potential applications, we implement an XOR gate based on these structures. In conclusion, our work provides a novel approach for on-chip OAM manipulation, with significant potential in optical computing and related fields.

Supplementary material of

# On-chip processing of optical orbital angular momentum

*Charles Chen[1]*

*1 State Key Laboratory for Mesoscopic Physics & Department of Physics, Collaborative Innovation Center of Quantum Matter & Frontiers Science Center for Nano-optoelectronics, Peking University, Beijing 100871, P. R. China*

# 1. Decomposition of Laguerre-Gaussian modes

The Laguerre-Gaussian (LG) mode is a very classic orbital angular momentum (OAM) mode, and its electric field expression contains the Laguerre polynomial term[1]. Previous works have demonstrated that, based on the relationship between the Laguerre polynomial and the Hermite polynomial[2, 3], the LG mode can be decomposed. Fig. S1 shows the decomposition of the LG mode summarized by us. Here, HG represents the Hermite-Gaussian (HG) mode. The "o" and "e" in $LG_{0i}^{o}$ and $LG_{0i}^{e}$ denote the even and odd symmetries in the azimuthal direction.

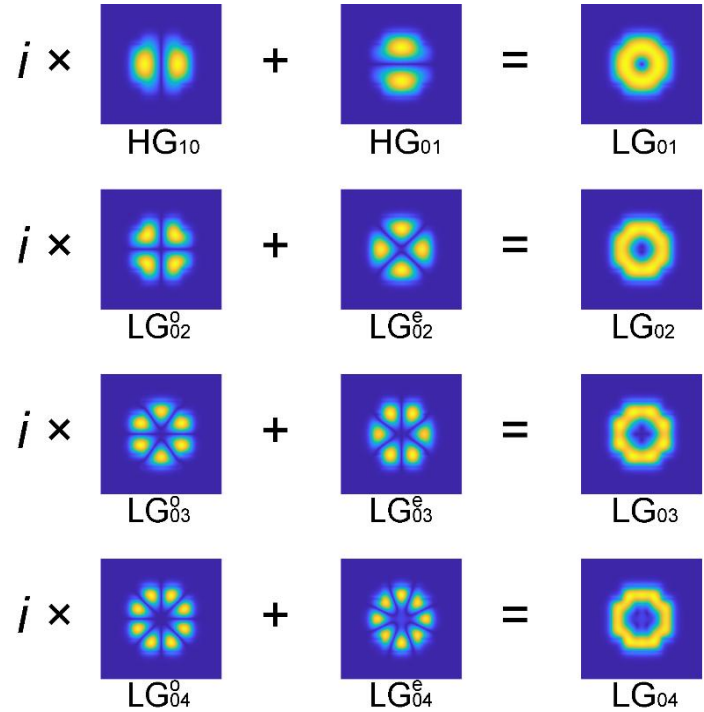

**Fig. S1:** Decomposition of LG modes. The $LG_{01}$, $LG_{02}$, $LG_{03}$ and $LG_{04}$ modes OAM, and they are composed of two corresponding modes with a $\frac{\pi}{2}$ phase difference.

## 2. Coupling between OAM$_{l=3}$ (OAM$_{l=4}$) mode and other OAM modes

Cross-shaped waveguide (CSW) structures can support OAM modes with different topological charge *l*. OAM$_{l=1}$ mode in CSW-1, OAM$_{l=2}$ mode in CSW-2, OAM$_{l=3}$ mode in CSW-3 and OAM$_{l=4}$ mode in CSW-4 are momentum-matched, so each two of them can couple with each other.

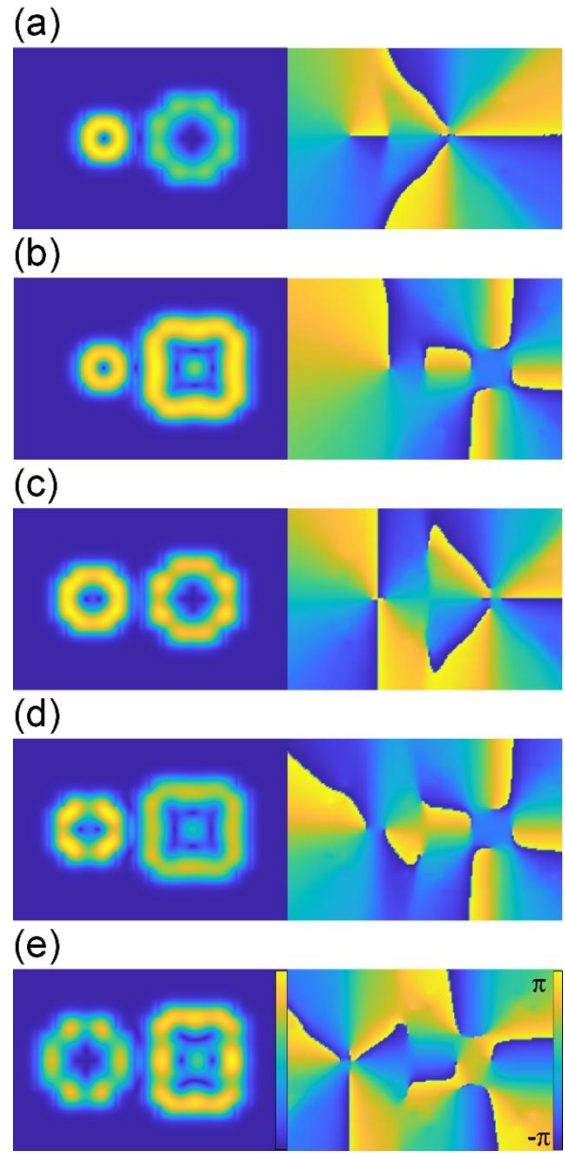

**Fig. S2:** Coupling situations when one of the two modes is OAM$_{l=3}$ or OAM$_{l=4}$ mode; (a-e) Intensity and phase distributions of coupling between (a) OAM$_{l=1}$ and OAM$_{l=3}$; (b) OAM$_{l=1}$ and OAM$_{l=4}$; (c) OAM$_{l=2}$ and OAM$_{l=3}$; (d) OAM$_{l=2}$ and OAM$_{l=4}$; (e) OAM$_{l=3}$ and OAM$_{l=4}$.

## 3. Fabrication process of CSW

To fabricate the CSW structure, processes such as electron beam lithography (EBL), reactive ion etching (RIE), and inductively coupled plasma (ICP) are required to perform multiple etching and coating operations on the substrate ($SiO_2$ on Si). The construction process of the sample is shown in Fig. S3[2].

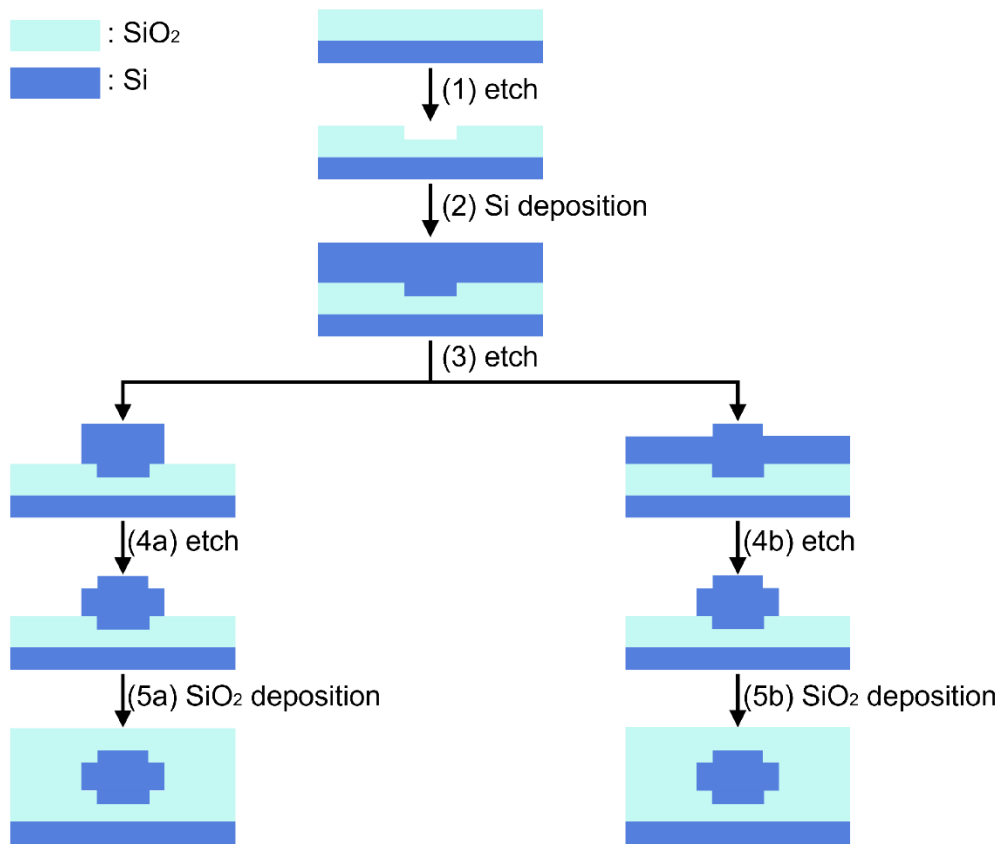

**Fig. S3:** Schematic diagram of CSW construction process. Step (4a), (5a) and step (4b), (5b) are two different but possible approaches to fabricate the sample.